\documentclass[10pt,twocolumn]{article}
\usepackage[letterpaper, top=0.5in, bottom=0.5in, left=0.5in, right=0.5in, columnsep=0.25in]{geometry}
\usepackage{amssymb}
\usepackage{amsmath}
\usepackage{xspace}
\usepackage{graphicx}
\usepackage{multirow}
\usepackage{multicol}
\usepackage{booktabs}
\usepackage[skip=4pt]{caption}
\usepackage{hyperref}
\usepackage{listings}
\usepackage{xcolor}
\usepackage{titlesec}
\usepackage{tabularx}
\usepackage{makecell}
\usepackage{subcaption}
\usepackage[normalem]{ulem}
\usepackage{float}
\usepackage[nameinlink]{cleveref}
\usepackage{mathtools}
\usepackage[all]{nowidow}
\newcommand{\sysname}{\textsc{ShellSieve}\xspace}
\newcommand{\secref}[1]{\S \ref{#1}\xspace}

\newenvironment{packeditemize}{
\begin{list}{$\bullet$}{
\setlength{\itemsep}{1.5pt}
\setlength{\labelwidth}{8pt}
\setlength{\leftmargin}{10pt}
\setlength{\labelsep}{3pt}
\setlength{\listparindent}{\parindent}
\setlength{\parsep}{1.5pt}
\setlength{\parskip}{1.5pt}
\setlength{\topsep}{1.5pt}}}{\end{list}}
\newcommand{\codespan}[1]{\texttt{\small#1}\xspace}
\newcommand{\codespanit}[1]{\texttt{\small#1}\xspace}

\newcolumntype{R}{>{\raggedleft\arraybackslash}X}
\newcolumntype{H}{>{\hsize=0.5\hsize\raggedleft\arraybackslash}X}

\titleformat{\paragraph}[runin]{\bfseries}{}{0pt}{}
\titlespacing*{\paragraph}{0px}{0.8ex plus 0.2ex minus 0.2ex}{0.5em}
\titlespacing*{\section}{0pt}{1.4ex plus 0.4ex minus .2ex}{1.2ex plus .4ex minus .1ex}
\titlespacing*{\subsection}{0pt}{1.2ex plus 0.3ex minus .1ex}{1.1ex plus .4ex minus .1ex}
\newfloat{listing}{htbp}{lop}
\floatname{listing}{Listing}

\lstdefinestyle{tiny}{
    basicstyle=\tiny,
}

\newtheorem{definition}{Definition}
\crefname{definition}{Definition}{Definitions}

\newcommand{\blk}{\mathrel{\vartriangleleft}}              
\newcommand{\blks}{\mathrel{\vartriangleleft^{\mkern1mu+}}} 
\newcommand{\blkw}{\mathrel{\vartriangleleft^{\mkern1mu-}}} 

\newcommand{\myhead}[1]{\makebox[0pt][r]{\rotatebox{-30}{\textbf{#1}}}}

\begin{document}
\title{One Goal, Many Commands: Characterizing Denylist Fragility in AI Agents}

\author{
  Chuyang Chen \\
  \textit{The Ohio State University} \\
  \textit{chen.13875@osu.edu}
  \and
  Zhiqiang Lin \\
  \textit{The Ohio State University} \\
  \textit{zlin@cse.ohio-state.edu}
}
\date{}
\maketitle


\begin{abstract}
    The adoption of AI agents is increasing rapidly. Terminal AI agents, i.e., AI agents that run in terminal environments, are a widely used type of AI agents. Terminal AI agents rely heavily on shell command execution to interact with the host systems. They adopt a three-list command-gating mechanism to mitigate security risks introduced by command execution, with denylists serving as the load-bearing component. However, modern operating systems often ship a large, ever-expanding set of shell commands with complex functionalities. Our observation is that even a built-in denylist of Claude Code, well-maintained by its developers, can overlook bypass commands that invalidate its effectiveness. Such negligence leads to fragile command denylists that cannot even block operations that practitioners expect them to block. \looseness=-1

    This paper presents the first systematic characterization of command denylist fragility in terminal AI agents. The paper formalizes the command denylist fragility problem and proposes an LLM-driven pipeline, \sysname, to detect such fragility. It prompts the LLM to propose possible bypasses and iteratively repairs them using feedback from a validator that executes them in a sandbox. In the evaluation, we applied \sysname to 1,709 real-world command denylists (containing 13,332 denylist rules) collected from GitHub. The evaluation shows several key findings, including that 69.0--98.6\% of the denylists are fragile, that this fragility occurs consistently across projects and agents, and the validity of several possible root causes for this fragility. Our pipeline and findings will hopefully facilitate future research and practice regarding the command denylists used by AI agents. \looseness=-1
\end{abstract}

\section{Introduction}
\label{sec:intro}

AI agents running in terminal environments, such as Claude Code and Codex, have recently emerged as versatile personal assistants, handling tasks ranging from software development to tax filing on behalf of human users~\cite{merrill2026terminal-bench}. One of the most important and yet most dangerous capabilities of these \textit{terminal AI agents} is their ability to dynamically synthesize and execute shell commands to fulfill context-dependent needs of the tasks~\cite{liu2026your}. Such an ability creates a \textit{dilemma between flexibility and security}: the agents are useful because they can choose and use various commands on the host system with broad autonomy and minimal human intervention, while security may be compromised by the extensive permissions required by such ``freedom.''\looseness=-1 

Currently, the common practice to mitigate the dilemma is to gate commands using three lists: a denylist to filter out obviously dangerous operations, an allowlist to permit obviously benign operations, and an implicitly defined (though sometimes customizable) ask-list that contains everything else and sending them to an evaluator (users themselves for manual audit, or an LLM-based judge in some designs) to get a final decision. Two facts make the denylist the load-bearing component in this \textit{three-list command-gating mechanism}: First, commands in the denylist are inherently dangerous and should not be executed under any circumstances (according to the semantics of ``denylists''). The aftermath of these commands should be deterministic and severe. Second, \textit{approval fatigue} causes the users to approve 93\% of command execution requests~\cite{claude}. Thus, the ask-list largely regresses to an allowlist, significantly compromising its effectiveness. \looseness=-1

\paragraph{The problem.} However, denylists can be fragile, as \textit{the same goal (e.g., performing a harmful operation) can often be achieved by many commands}. Hurriedly compiled denylists may block some commands that perform the operations the authors want to forbid, but overlook others that bypass their protection. This situation worsens further, given that modern operating systems are essentially ``open worlds'' that often ship a large, ever-expanding set of shell commands to meet diverse development needs. The functionality of these commands is often beyond the knowledge of ordinary practitioners. Our observations (detailed in \secref{sec:observation}) on the built-in denylist in Claude Code's auto-mode and an arbitrarily sampled command denylist from GitHub confirm this intuition: even the denylist shipped with a widely used agent by default and well-maintained by its developers can overlook bypass commands that compromise its protection, let alone the denylists compiled by regular users of these agents. \looseness=-1

Therefore, a critical problem requiring urgent attention is to evaluate and characterize this fragility to facilitate future research and practice. \looseness=-1

\paragraph{Challenges and methodology.} Two obstacles stand in the way of the characterization. First, it is hard to propose candidate operations that a denylist aims to block, and to identify bypass commands that can perform them. Second, even with candidate operations and bypass commands, it is hard to validate them, since their execution is ephemeral. \looseness=-1

To address the challenges, we propose \sysname, an LLM-driven pipeline with two targeted designs for characterization. First, \sysname leverages an LLM to propose candidate operations and bypass commands. LLMs' expert-level capability to synthesize shell commands~\cite{sladic2024llm} enables them to produce thousands of diverse, high-quality candidates that no traditional technique can propose. Second, \sysname defines and validates candidate operations and bypass commands based on the side effects they leave in a sandbox environment after execution. Thus, it avoids ambiguity and false positives in the characterization. \looseness=-1

\paragraph{Findings.} We evaluate 1,731 real-world AI agent denylists (containing 13,332 denylist rules) collected from GitHub with \sysname and reveal several key findings:\looseness=-1
\begin{packeditemize}
    \item First, the denylist fragility problem is severe. 69.0--98.6\% denylists cannot fully block operations they aim to block.\looseness=-1
    \item Second, the severity of the fragile denylist problem is consistently high across projects and agents. The proportion of denylists that overlook bypass commands is consistently above 44.4\% across projects that receive varying levels of attention. Similarly, the proportion is always higher than 46.2\% for projects that use different agents.\looseness=-1
    \item Third, we statistically confirmed two root causes of the denylist fragility problem. One is that the authors of denylists ignore less-known commands, creating bypasses. Another is that multi-purpose commands are deliberately unlisted from the denylists for their benign use. Still, they may become bypasses when used in ways that perform the operations the denylists aim to block.\looseness=-1
    \item Fourth, fixing fragile denylists will require adding numerous commands (averaging 217 if they aim to block specific operations) to complete them, which is a huge burden for anyone who wants to resolve the problem.\looseness=-1
\end{packeditemize}

\paragraph{Contributions} Overall, we make three contributions: \looseness=-1
\begin{packeditemize}
  \item We identify and formalize the command denylist fragility problem for terminal AI agents. We argue that despite their load-bearing role in the command gating mechanism of AI agents, command denylists can be fragile and overlook bypasses that compromise their effectiveness. \looseness=-1
  
  \item We design and implement \sysname, an LLM-driven pipeline, to automatically discover bypass commands that can perform operations a denylist aims to block. It enables large-scale characterization of the problem across corpora comprising thousands of denylists.\looseness=-1

  \item We collect and construct a dataset of 1,709 denylists (containing 13,332 denylist rules) and evaluate them using \sysname. The evaluation reveals several key findings that may bring insights and directions for future research and practice. \looseness=-1
  
  
\end{packeditemize}

The code and data will be released upon publication. \looseness=-1

\section{Background}
\label{sec:background}

\begin{figure}[t]
  \centering
  \includegraphics{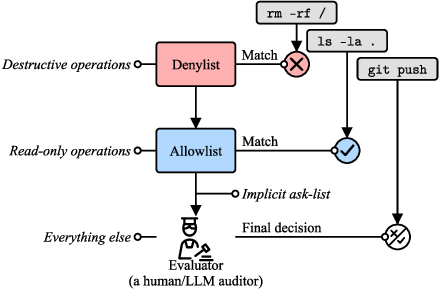}
  \caption{A three-list command-gating model. Implementations may vary in some details.}
  \label{fig:2-three_lists}
\end{figure}

\paragraph{Terminal AI agents.} AI agents are software driven by AI models that interact with host systems to perform various tasks on behalf of human users. Terminal AI agents are a type of AI agent that run in terminal environments and use shell commands to interact with the system and have been deployed and used in the real world~\cite{merrill2026terminal-bench, cheng2026terminal-world}. Some well-known terminal AI agents include Claude Code, a terminal-based general-purpose assistant from Anthropic~\cite{claude}, Codex, a coding agent from OpenAI~\cite{codex}, and OpenCode, an open-source, community-driven agent compatible with different model providers~\cite{opencode}. All these agents share the same architecture: an LLM driver, a harness layer that orchestrates requests to and responses from the LLM, and various tools, including the shell command execution tool mentioned above. However, the ability to execute shell commands introduces a dilemma between flexibility and security. The agents are useful because they can run whatever commands a task requires. Yet this same freedom allows an adversary who hijacks it to perform arbitrary, harmful operations on the host. \looseness=-1

\begin{listing}[t]
    \centering
    \begin{lstlisting}
"deny": [
    "cat *.env*",
    "head *.env*",
    "tail *.env*",
    "less *.env*",
    "more *.env*",
    "grep *.env*"
]
    \end{lstlisting}
    \caption{Running example, a Claude Code command denylist where ``*'' is a wildcard that matches any string, including the empty string. \looseness=-1}
    \label{lst:2-running_example}
\end{listing}

\paragraph{Three-list command gating.} The common practice to mitigate the dilemma is to gate the shell commands before execution using three lists, as illustrated in \autoref{fig:2-three_lists}:
\begin{packeditemize}
  \item \textbf{Denylist.} Denylists block obviously dangerous operations that the agent must never run. Commands in denylists typically conduct destructive operations (such as file deletion) or pose an obvious threat (such as secret leaks). \autoref{lst:2-running_example} gives an example denylist in a GitHub project that uses Claude Code. The example will be revisited in \secref{sec:2_1-custom_bypass}. \looseness=-1
  \item \textbf{Allowlist.} Allowlists permit benign operations that the agent may run without any intervention, such as reading a non-confidential file. \looseness=-1
  \item \textbf{Ask-list.} Everything not matched by the allowlist or the denylist goes into the ask-list and is delegated to an authoritative evaluator, typically the user, for a final decision~\cite{configure, rules, 2026permissions, policy}. Recently, more agents are exploring using an LLM-based judge as an evaluator~\cite{claude}. \looseness=-1
\end{packeditemize}
This basic three-list command-gating architecture is observed in various terminal AI agents. Several notes should accompany the model. First, the denylist takes the highest precedence (thus it is at the top in \autoref{fig:2-three_lists}), meaning that any commands it contains will be rejected directly, neither matched by the allowlist nor checked by the evaluator. Second, the denylist and allowlist are typically customizable by users to meet their specific needs and can sometimes be updated while the agent is running to take commands from the ask-list, avoiding tedious, repeated approval requests. Third, the architecture described in this section is an abstract model, and implementations across different agents may vary in some details, e.g., Claude Code includes an extra built-in denylist in its auto-mode. \looseness=-1

\paragraph{Approval fatigue.} The ask-list is advertised as a fine-grained safety net, neither too permissive nor too restrictive, whereas the promise is largely compromised by approval fatigue. Users should, by expectation, carefully audit commands proposed to run by the ask-list. However, real-world data shows that the expected approval process breaks: users accept 93\% of approval requests~\cite{claude}. The overwhelming majority of decisions means users tend to approve anything~\cite{claude, reeves2021encouraging} unthinkingly. The ask-list largely regresses into an allowlist, and the boundary it should enforce thus loosens. An LLM judge, as the evaluator, will not suffer from approval fatigue. However, it may encounter LLM-specific issues such as hallucinations~\cite{chrysos2025agentic} and adversarial suffices~\cite{andriushchenko2025jailbreaking, zou2023universal} though. Therefore, denylists become the load-bearing component in the three-list command-gating mechanism. \looseness=-1


\section{Observations}
\label{sec:observation}

This section presents our observations on two denylists to demonstrate the prevalence of fragile command denylists. In \secref{sec:2_1-awk-bypass}, we run through the built-in denylist in Claude Code's auto-mode (mentioned in \secref{sec:background}) to show that even a denylist shipped with a widely used agent by default and well maintained by its developers can overlook bypasses. In \secref{sec:2_1-custom_bypass}, we examine the denylist of an open-source project on GitHub to show that the same issue occurs in denylists compiled by regular users. \looseness=-1

\subsection{Bypass of the Claude Code Built-in Denylist}
\label{sec:2_1-awk-bypass}

\paragraph{Purpose of the denylist.} The built-in denylist in Claude Code's auto-mode enumerates code-execution commands, including language interpreters, package managers' script running subcommands, local and remote shells, and forces them through the LLM judge, forbidding the customizable allowlist from silently approving them, as these commands can conduct arbitrary operations on the host system, and thus shouldn't be mindlessly waved through~\cite{claude}. \autoref{tab:2_1-code_exec} lists the commands in this denylist, confirmed by experiments. \looseness=-1

\paragraph{The bypass.} However, the enumeration is still incomplete. It overlooks commands that can also execute arbitrary code and are shipped in all commonly used Linux distributions (which are host systems that Claude Code often lives in). One of them is AWK, a POSIX-mandated utility present on every Unix and Unix-like host. Though the primary use of AWK is to process texts, the domain-specific language (DSL) used by AWK is Turing-complete, and more importantly, AWK can invoke Bash and pass arbitrary Bash scripts in the invocation. The fact that AWK is not in the denylist essentially bypasses the rule that covers Bash. Experiments confirmed that AWK can bypass the denylist in Claude Code's auto-mode, and a disclosure has been sent to Anthropic, the company that develops Claude Code. \looseness=-1

\paragraph{The observation.} This bypass confirms the observation that denylists can be fragile, even when well-maintained by the agent developers, as they may overlook bypass commands. \looseness=-1

\subsection{Bypasses of a Customized Denylist}
\label{sec:2_1-custom_bypass}


The above denylist is a single special example. It indicates how fragile command denylists can be, though. Now, we demonstrate a real denylist for an open-source project on GitHub, compiled by the project's developers, as an example of fragility in user-customized command denylists deployed more widely across projects that use these agents. \looseness=-1

\paragraph{Purpose of the denylist.} \autoref{lst:2-running_example} shows the denylist. It is written in Claude Code's denylist format, where the wildcard \codespan{*} matches any string, including the empty string. Clearly, all commands blocked the denylist read files that define environment variables and print their values to stdout, which is exactly the operation that the denylist targets. \looseness=-1

\paragraph{The bypasses.} Many commands can perform the same operation, and those blocked by this denylist are only a small proportion among them. Examples include \codespan{cp .env /dev/stdout} and \codespan{diff .env /dev/null}. The denylist considers only the most common commands for this purpose, thereby overlooking others. \looseness=-1

\paragraph{The observation.} These bypasses confirm the observation that agent users adopt fragile denylists in the wild. \looseness=-1

\begin{table}[t]
    \centering
    \begin{tabularx}{\linewidth}{Xr}
        \toprule
        \textbf{Category} & \textbf{Commands} \\
        \midrule
        \multirow[t]{2}{\hsize}{Interpreters} & \texttt{python}, \texttt{python3}, \texttt{python2}, \texttt{node}, \\
        & \texttt{deno}, \texttt{tsx}, \texttt{ruby}, \texttt{perl}, \texttt{php}, \texttt{lua} \\
        \multirow[t]{2}{\hsize}{Package managers} & \texttt{npx}, \texttt{bunx}, \texttt{npm run}, \texttt{yarn run}, \\
        & \texttt{pnpm run}, \texttt{bun run} \\
        Local and remote shells & \texttt{bash}, \texttt{sh}, \texttt{ssh} \\
        \bottomrule
    \end{tabularx}
    \caption{The code execution commands in the built-in denylist of Claude Code's auto-mode.}
    \label{tab:2_1-code_exec}
\end{table}

\section{Threat Model}
\label{sec:3-threat}

\paragraph{Attack scenarios.} We consider an adversary whose objective is to cause the agent to execute a command that produces an attacker-chosen, security-relevant effect on the host, such as exfiltrating credentials or irreversibly destroying data. The adversary, however, does not hold a shell on the host directly, as the gating mechanism would be moot in that case. We further assume the adversary knows the gating policy in full, including the denylist. It is a realistic Kerckhoffs-style assumption, given that these lists are often managed in a per-project setting together with the source code in a Git repository~\cite{configure, rules, 2026permissions}. The adversary should be able to dictate the agent's proposed commands. This premise reflects the fact that AI agents routinely ingest untrusted content that may carry adversarial payloads to conduct \textit{indirect prompt injection} as part of normal operation~\cite{shi2025prompt, zhan2024injecagent, khan2026clinejection}. Our analysis does not depend on any single injection technique. It requires only the now-standard observation that the commands reaching the gate cannot be assumed benign~\cite{liu2026your}. \looseness=-1

\paragraph{Trust boundary.} The trust boundary in our model lies precisely at the command-gating layer. Everything upstream, including the LLM driver, the harness, and any content they have consumed, is untrusted, since a compromised alignment places the proposed command entirely under adversarial control. Everything downstream, including the shell, the binaries it can reach, and the host on which they run, is trusted only in the operational sense that it faithfully executes whatever the gate admits, performing no further check of its own. The gate with denylists like those shown in \S\ref{sec:background} and \ref{sec:observation} is thus the sole point that can stop the adversarial operations.\looseness=-1

\paragraph{Out-of-scope problems.} Some adjacent threats fall outside our scope, including how an adversary compromises the model's alignment in the first place, which is taken as given, and implementation-level defects in the gating layer itself, as our argument is strictly stronger without them: we show that the denylist fails even when it is implemented and matched flawlessly. Attacks that bypass the shell-execution tool entirely, such as reverse-shell attacks~\cite {command}, are not considered either, as they are orthogonal to the command-gating mechanism explained above. \looseness=-1
\section{Problem Fomalization}
\label{sec:problem}

Building on the previous observations, this section formalizes the problem of fragile command denylists. As mentioned in \secref{sec:intro}, the formalization should enable us to identify and observe what a denylist aims to block. Definitions in this section fix the objective for the rest of this paper, which will refer to these formal definitions as guidance when implementing our system.\looseness=-1

\subsection{Denylists and deny-rules}
\label{sec:problem-denylist}

We begin by providing, in \cref{def:4-denylists}, the formal definitions of denylists and denylist rules (or deny-rules for short), the elements that form them. Later explanations use \autoref{lst:2-running_example} instead of \autoref{tab:2_1-code_exec} as the running example because the auto-mode denylist is hardcoded in the private code of Claude Code (the table was discovered via experiments), and no detail is provided to validate our definitions. \looseness=-1

\begin{definition}[Deny-rules and denylists]
    \label{def:4-denylists}
    A \textit{deny-rule} $r$ is a predicate over a command with its argument list:
    \[
        r: C\times A^n \rightarrow \{\top, \bot\}
    \]
    where $C$ is the set of all available commands, $A$ is the set of all possible argument strings, and $n$ is the number of command arguments. A command $c$ with argument list $a$ is blocked by $r$, denoted as $c\blk_a r$, if and only if
    \[
        r\left(c, a\right) = \top
    \]
    A \textit{denylist} $L$ is a set of deny-rules:
    \[
        L\in \mathcal{P}(R)
    \]
    where $R$ is the set of all possible deny-rules, and $\mathcal{P}(\cdot)$ denotes power sets. A command $c$ with argument list $a$ is blocked by $L$, denoted as $c\blk_{a} L$, if
    \[
        \bigvee_{r\in L} r\left(c, a\right) = \top
    \]
    i.e., a command is blocked by a denylist if any deny-rule in the list blocks it.
\end{definition}
The internal form of denylists and deny-rules is deliberately abstracted, so that the definitions that follow are implementation-agnostic and apply uniformly across agents. The only capability a deny-rule must expose is a decidable match verdict on an invocation. How that verdict is computed (e.g., either by using prefix matching~\cite{codex} or glob patterns~\cite{opencode}) is immaterial to our analysis. In \autoref{lst:2-running_example}, each line between the brackets is a deny-rule. The pattern \codespan{cat *.env*} matches the command \codespan{cat project.env}. Therefore, the command should be blocked if it is the one undergoing command gating. The six deny-rules in this listing form a denylist together.

Several notes about the definitions are worth mentioning: First, matching is a property of the surface form of a command invocation, not of its effect on the host. A deny-rule fires on what a command and its arguments look like, whereas the operations an invocation achieves are settled only when it runs (\secref{sec:problem-goals}). Second, a denylist is a purely negative specification. It enumerates invocations to reject, so any command invocation outside the union of its rules is admitted by default, and its protection is bounded by whatever its authors thought to enumerate.

\subsection{Blocked Operations}
\label{sec:problem-goals}

Before defining what a denylist aims to block, we first define what a denylist actually blocks, which should, intuitively, be the operations that the blocked commands perform. Take \autoref{lst:2-running_example} for example. All six deny-rules block commands that print the contents of files with env-suffixed names. Thus, this denylist should aim to block ``printing the contents of files with env-suffixed names.'' \cref{def:4-op} defines operations and blocked operations.

\begin{definition}[Operations and blocked operations]
    \label{def:4-op}
    Given an aspect (e.g., the file system or network layers) of a host system $h$, an \textit{operation} $o$ is a function over the states of that aspect:
    \[
        o: S \rightarrow S
    \]
    Here, $S$ is the set of all possible states of that aspect, and $p$ maps an old state to a new state. $p$ is an operation performed by a command $c$ with argument list $a$ on $h$, denoted as $c \vdash_{h, a} p$, if and only if the states before and after execution $s$ and $s'$ satisfy 
    \[s' = o(s)\] 
    The set of all operations performed by $c$ on $h$ is denoted as $\mathcal{O}_h(c)$. In turn, $o$ is an operation \textit{blocked} by denylist $L$ on $h$, denoted as $o \blk_h L$, iff.
    \[
        \exists c \in C_h, a \in A^n. c\vdash_{h, a} o \wedge c\blk_a L
    \]
    where $C_h$ is the set of all available commands on $h$, and $A$ is the set of all possible argument strings. The set of all operations blocked by $L$ on $h$ is denoted as $\mathcal{O}_h(L)$.
\end{definition}
Instead of ephemeral behaviors, the definition focuses on the side effects that a command leaves in the host system after execution, and scopes the characterization of commands to the essential ``aftermath'' they cause. As an example, \codespan{cat project.env} and \codespan{cp project.env /dev/stdout} invoke different syscalls. The former writes the file content from a buffer directly to the special file descriptor the system maintains for stdout after reading it, while the latter creates a new file descriptor for \codespan{/dev/stdout} and writes to it. However, they leave the system in the same post-execution state where the inode of stdout holds the content read from \codespan{project.env}. These two commands perform the same operation. \looseness=-1

Specifically, this paper restricts the blocked operations under investigation into seven kinds of file system operations, each represented by an ID and an argument list, as shown in \autoref{tab:4-scopes}. File system operations are a major type of operation that malware performs~\cite {bayer2009view} and are easy to record and analyze. Other operations, such as network accesses, will require complex simulations, e.g., fake HTTP servers, in the experiments and are thus not considered in this paper. Future work may further investigate the other operations. The seven kinds of operations are explained as follows:
\begin{packeditemize}
    \item \textbf{\codespan{op\_read}.} The content of the file at \codespanit{path} is read and printed to stdout or stderr. \looseness=-1
    \item \textbf{\codespan{op\_write}.} \codespanit{content} is written to the file at \codespanit{path}. \looseness=-1
    \item \textbf{\codespan{op\_create}.} A new file is created at \codespanit{path}. \looseness=-1
    \item \textbf{\codespan{op\_delete}.} The file at \codespanit{path} is deleted. \looseness=-1
    \item \textbf{\codespan{op\_copy}.} The file at \codespanit{path1} is copied to \codespanit{path2}. \looseness=-1
    \item \textbf{\codespan{op\_chmode}.} The mode of the file/directory at \codespanit{path} is changed to \codespanit{mode}. This is typically conducted by the \codespan{chmod} command on Linux. \looseness=-1
    \item \textbf{\codespan{op\_chowgr}.} The owner/group of the file/directory at \codespanit{path} is changed to \codespanit{owner\_and\_group}. This is typically conducted by the \codespan{chown} command on Linux. \looseness=-1
\end{packeditemize}

As an example, all seven rules in \autoref{lst:2-running_example} correspond to a set of blocked operations that read the \codespan{.env} files: \codespan{op\_read(".env")}, \codespan{op\_read("project.env")}, ... In our implementation, arguments of blocked operations have placeholders to denote an infinite set of operations following the same pattern (see \secref{sec:5-method_overview}). \looseness=-1

The seven kinds cover three components of a host's file system that can be modified: file contents, file paths, and file metadata. Several notes are worth mentioning here: First, the list does not include a kind for ``code execution'', because ``execution'' is not a single side effect that can be observed in the host's file system. Instead, ``execution'' can indirectly leave other side effects, such as file creation or deletion. Thus, ``execution'' is reasonably excluded from being considered as one single operation kind. Denylists that aim to block code execution, like the one in \autoref{tab:2_1-code_exec}, will be categorized as targeting multiple concrete file system operations, such as file deletion or creation. Second, the list is designed to be as minimal as possible. Thus, it does not include ``moving a file,'' which can be decomposed into a copy and a deletion (of the file at the original path). Third, the list does not include some less-used metainformation, such as xattr~\cite{xattr7}, to focus on the most common operations first. \looseness=-1

Then, \textit{the blocked operations $\mathcal{B}(L)$ are used as an approximation of the operations a denylist $L$ aims to block,} considering that a precise depiction, essentially equivalent to the intents of the authors of the denylist, would require an in-depth interview to investigate. We argue that this is a good enough approximation based on two intuitions:
\begin{packeditemize}
    \item First, though the authors of a denylist sometimes forget to include some commands, they should not be expected to leave an operation they want to block \textit{completely untouched}, which means that there should be at least one command that performs the operation in the denylist. For example, the authors of the denylist in \autoref{tab:4-scopes} forget that \codespan{cp} and \codespan{diff} can also print the content of the \codespan{.env} files, but they do block some obvious commands (\codespan{cat}, \codespan{head}, ...) for this purpose. \looseness=-1
    \item Second, authors tend to minimize the denylist, as an unnecessarily over-general denylist will block too many benign operations and hinder the normal development process, violating the design purpose of denylists in the three-list gating mechanism. \looseness=-1
\end{packeditemize}

\begin{table}[t]
    \centering
    \begin{tabularx}{\linewidth}{Xr}
    \toprule
           \textbf{Operation ID} & \textbf{Arguments} \\
    \midrule
         \texttt{op\_read} & \codespanit{path} \\
         \texttt{op\_write} & \codespanit{path}, \codespanit{content} \\
        \texttt{op\_create} & \codespanit{path} \\
         \texttt{op\_delete} & \codespanit{path} \\
         \texttt{op\_copy} & \codespanit{path1}, \codespanit{path2} \\ 
         \texttt{op\_chmode} & \codespanit{path}, \codespanit{mode} \\
         \texttt{op\_chowgr} & \codespanit{path}, \codespanit{owner\_and\_group} \\
    \bottomrule
    \end{tabularx}
    \caption{Blocked operations investigated in this paper.\looseness=-1}
    \label{tab:4-scopes}
\end{table}

\subsection{Incompletely Blocked Operations}
\label{sec:problem-alternatives}

The previous subsection defines blocked operations to characterize what a denylist aims to block. This section will investigate whether a denylist can fully block an operation. Specifically, \cref{def:4-full} defines full blocked and incompletely blocked operations. \looseness=-1

\begin{definition}[Fully and incompletely blocked operations]
    \label{def:4-full}
    Given a denylist $L$ and an operation $o$ on a host system $h$, $o$ is fully blocked by $L$ on $h$, denoted as $o \blks_h L$, if and only if
    \[
        o\blk_h L \wedge \forall c\in C_h, a\in A^n. c\vdash_{h, a} o \implies \neg c\blk_a L
    \]
    where $C_h$ is the set of all commands provided by $h$, and $A$ is the set of all possible argument strings. $p \blks_h L$ essentially means that every command on $h$ that may perform $o$ has been blocked by $L$. In turn, $o$ is incompletely blocked by $L$ on $h$, denoted as $o \blkw_h L$, if and only if,
    \[
        o\blk_h L \wedge \exists c\in C_h, a\in A^n. c\vdash_{h, a} o \wedge \neg c\blk_a L
    \]
    The set of all operations fully/incompletely blocked by $L$ is denoted as $\mathcal{O}_h^+(L)$/$\mathcal{O}_h^-(L)$. Note that a blocked operation, as defined in \cref{def:4-op} ($o \blk_h L$), is either fully blocked ($o \blk_h^+ L$) or incompletely blocked ($o\blk_h^- L$), and thus $\mathcal{O}_h^+(L)\cap \mathcal{O}_h^-(L) = \emptyset$ and $\mathcal{O}_h^+(L)\cup \mathcal{O}_h^-(L) = \mathcal{O}_h(L)$. Any command $c$ not in $L$ that perform an operation $o\in \mathcal{O}_h^-(L)$ is called a \textit{bypass command} or \textit{bypass} in short, denoted as $c \succ_{h, o} L$. The set of all bypasses of $L$ that perform $o$ on $h$ is denoted as $\mathcal{B}_{h, o}(L)$. \looseness=-1
\end{definition}
A denylist that incompletely blocks some operations provides no effective protection, as an attacker can easily perform them using bypass commands. Take \autoref{lst:2-running_example} as an example. The operation \codespan{op\_read(".env")} is incompletely blocked, as the command \codespan{diff .env /dev/null} can also read and print the content of \codespan{.env} while not being blocked. Thus, the \codespan{diff} command is a bypass. \looseness=-1

Several notes are worth mentioning. First, the relations $o \blks_h L$ and $o \blkw_h L$ are not symmetric. $o \blkw_h L$ is not a negation of $o \blks_h L$. It has an extra precondition $o\blk L$, which excludes from consideration those operations completely ignored by the denylist, because the authors likely keep these operations untouched on purpose, for reasons such as that the operations are trivial for their projects. Accounting for operations that do not meet the precondition will result in a scope that is too large and contains excessive noise. For example, \autoref{lst:2-running_example} does not cover the operation \codespan{op\_read("README")} at all, not because they forget it, but because reading this file does not imply any security issues. Second, bypass commands are not novel exploits and do not require vulnerabilities. They can be commands leveraged in their design usage, but for an attacker's purpose (see the \codespan{diff} command). Third, a bypass command does not have to be precisely equivalent to a command blocked by the denylist and only has to perform the same blocked operation. As explained before, the \codespan{cp} command has semantics completely different from \codespan{cat}, but it can perform the same operation respecting reading and printing file contents, which makes it essentially a bypass of the denylist for the blocked \codespan{op\_read} operations. \looseness=-1

\section{The \sysname Pipeline}
\label{sec:method}

Based on the formalization in the previous section, the core problem of characterizing denylist fragility is, given a denylist $L$ and an operation $o$ blocked by $L$ on a host system $h$ ($o\blk_h L$), to find whether bypasses exist (i.e., whether $\mathcal{B}_{h, o}(L) \neq \emptyset$) and what they are if they exist (i.e., the elements of $\mathcal{B}_{h, o}(L)$). This section explains the pipeline we use to compute $\mathcal{B}_{h, o}(L)$, The next subsection gives an overview (\secref{sec:5-method_overview}), followed by subsections explained the stages in the pipeline (\S\ref{sec:5-method_generation}--\ref{sec:5-method_match}), one for each.


\subsection{Overview}
\label{sec:5-method_overview}

\begin{table}[t]
    \centering
    \begin{tabularx}{\linewidth}{Xr}
    \toprule
        \textbf{Command with arguments} & \textbf{Operations} \\
    \midrule
        \codespan{cat \$N.env} & \codespan{op\_read(\$N.env)} \\
        \codespan{cp \$P1 \$P2} & \codespan{op\_copy(\$P1, \$P2)} \\
        \codespan{cp \$P /dev/stdout} & \codespan{op\_read(\$P)} \\
    \bottomrule
    \end{tabularx}
    \caption{Examples of the commands (input), arguments (output), and operations (output) consumed and produced by the pipeline. The arguments and operations can contain placeholders, including \codespan{\$N}for an arbitrary file name string, and \codespan{\$P},  \codespan{\$P1}, \codespan{\$P2} for arbitrary file paths. \looseness=-1}
    \label{tab:5-inout}
\end{table}

\paragraph{Input and output.} The inputs of our system are a denylist $L$ and a command $c$ provided by the host system $h$ that is not blocked by $L$ ($\neg c\blk_a L$ for some argument list $a$). Given an operation $o$, our system should decide whether $c\succ_{h, o} L$. If so, the command is a bypass, i.e., $c \in \mathcal{B}_{h, o}(L)$. For example, taking the denylist in \autoref{lst:2-running_example} and the command \codespan{cp} provided by the system, \sysname will output true if \codespan{cp} (the $c$) can read the content of a file \codespan{project.env} (reading file as the $o$) because it is not blocked by the denylist (the $L$) and thus a bypass. \looseness=-1

This is achieved by running a \textit{shared} pipeline in parallel on $L$ and $c$ to identify the operations they block/perform, i.e., $\mathcal{O}_h(L)$ and $\mathcal{O}_h(c)$, respectively. Then, the two sets of operations are compared, and if $\mathcal{O}_h(L) \cap \mathcal{O}_h(c) \neq \emptyset$, $c$ is a bypass of $L$ respecting the operations in $\mathcal{O}_h(L) \cap \mathcal{O}_h(c)$. For example, as shown in \autoref{tab:5-inout}, the pipeline will identify that $\mathcal{O}_h(L)$ is $\{\texttt{op\_read("\$N.env")}\}$ for $L$ being the denylist in \autoref{lst:2-running_example}, and $\mathcal{O}_h(c)$ is $\{\texttt{op\_read("\$P")}, \texttt{op\_copy(\$P1, \$P2)}\}$ for $c$ being the command \codespan{cp}. Then, $c \succ_{h, o} L$ for $o$ being \codespan{op\_read("\$N.env")} in $\mathcal{O}_h(L) \cap \mathcal{O}_h(c)$. Note that our implementation allows the arguments and operations to contain placeholders to represent infinite elements. \looseness=-1

\begin{figure*}[t]
    \centering
    \includegraphics{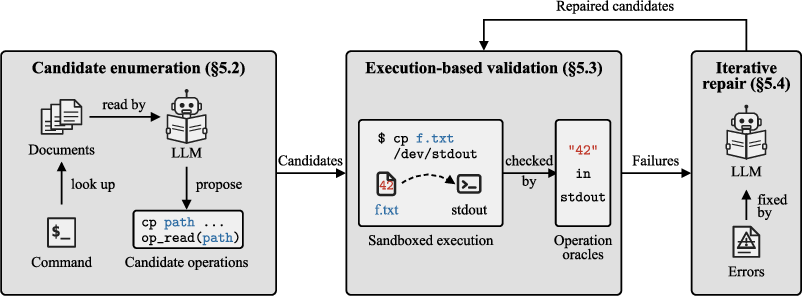}
    \caption{End-to-end \sysname pipeline. Each input (a command with arguments) flows through three stages. The \textit{candidate enumerator}, an LLM conditioned on \codespan{man} pages, \codespan{--help} outputs, and GTFOBins entries, proposes candidate operations that the command may perform. The \textit{validator} then executes the command against operation-specific oracles to decide whether these candidates are correct. Failures are routed through the \textit{repair loop} back to the LLM with diagnostic messages, for at most $T$ rounds. The confirmed operations are the pipeline's output.\looseness=-1}
    \label{fig:5-pipeline_overview}
\end{figure*}

\autoref{fig:5-pipeline_overview} shows an overview of the shared pipeline. The pipeline takes a command $c$ and outputs the operations it performs with the proper arguments. The operations are then aggregated over all arguments to compute $\mathcal{O}_h(c)$. The commands are retrieved from entries of a denylist $L$ (e.g., \codespan{cat} from \codespan{cat *.env*} in \autoref{lst:2-running_example}) or the executable paths (e.g., \codespan{cp} in \codespan{/usr/bin}) and go through the same three stages to produce the output, validated operations and arguments: \looseness=-1
\begin{packeditemize}
    \item \textbf{Candidate enumeration (\secref{sec:5-method_generation}).} Given the command, an LLM is prompted to propose a candidate operation within \autoref{tab:4-scopes} that the command may perform with proper arguments. Candidate operations and arguments that have obvious errors (e.g., wrong arguments) are rejected. Overlapping or identical operations are deduplicated. The LLM is re-prompted until there are enough candidates after rejection and deduplication. \looseness=-1   
    \item \textbf{Execution-based validation (\secref{sec:5-method_validation}).} Each command is executed in a sandbox with all placeholders in the arguments substituted by flag files and canary tokens, and the side effects it leaves are recorded. Specific oracles of the candidate operations check whether these side effects indicate that the operations are truly performed by the command. \looseness=-1
    \item \textbf{Iterative repair (\secref{sec:5-method_repair}).} Candidate operations that fail the validation are fed back to the LLM with error messages to get a repair. The repaired candidates then go through the validation process again. \looseness=-1
\end{packeditemize}
Ultimately, the operations performed by every command are compared with those blocked by the denylist to decide whether bypasses exist for that denylist (\secref{sec:5-method_match}).

\sysname{} relies on an LLM for candidate enumeration and iterative repair, the two stages that require understanding and synthesizing shell commands. This is a deliberate design choice. The inputs to these stages (man pages, \texttt{--help} output, and other documentation) are unstructured text, and the command-line interfaces they describe follow widely divergent conventions. Absent an LLM, one would have to assemble a pipeline from conventional techniques to recover the relationship between a command and its arguments: natural-language processing to mine these relations from documentation~\cite{wong2015dase, pandita2012inferring}, static analysis to supply structural information when source code is available~\cite{tan2007icomment}, and a constraint solver such as an SMT solver~\cite{monniaux2016survey} to find argument assignments that satisfy them. Such a pipeline is unattractive on several fronts: documentation- and specification-mining techniques are known to suffer high false-positive and false-negative rates~\cite{johnson2013why, amann2018systematic} and demand substantial engineering effort, while the static-analysis components face well-documented scalability limits~\cite{gosain2015static}. An LLM, by contrast, synthesizes shell invocations cheaply and at scale with little bespoke engineering~\cite{sladic2024llm}. Its principal weakness for our purpose is hallucination~\cite{ji2023mitigating, bang2025hallulens}, but this is precisely what the second stage neutralizes: every candidate is executed in a sandbox and checked against a predefined oracle, so unsound proposals are discarded rather than trusted. \looseness=-1

\subsection{Candidate Enumeration}
\label{sec:5-method_generation}

\begin{figure}[t]
    \centering
    \includegraphics[width=2.85in]{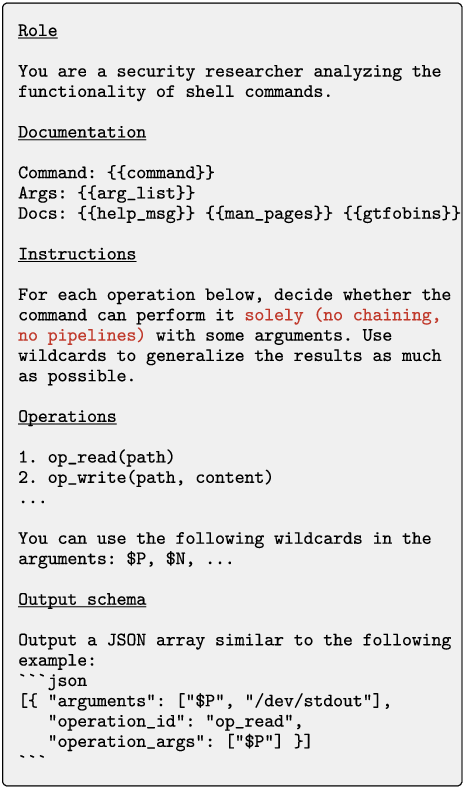}
     \caption{Abbreviated prompt template used in candidate enumeration. The LLM receives the command's documentation and is asked to propose a structured list of operation and argument candidates. Double braces mark the inserted contents.\looseness=-1}
    \label{fig:5-candidate_generation_prompt}
\end{figure}

During the candidate enumeration stage, an LLM proposes candidate operations and arguments. The command may perform the operations with the arguments. This is done by prompting the LLM with the command's documentation, including its help message, manual pages, and the corresponding entries on GTFOBins (a community-curated living-off-the-land attack database)~\cite{gtfobins} if any, to produce a structural list of candidate operations and arguments. The full prompt is shown in \autoref{fig:5-candidate_generation_prompt}. A caveat of the prompt is worth noting: we explicitly request that the reported operations be performed solely by this command, without chaining or pipes, to rule out cases in which the command itself does not meet the requirement but is concatenated with other commands that do.\looseness=-1

Obviously unsatisfactory candidates, such as those containing syntax errors, are discarded, and the LLM is re-prompted until it proposes a satisfactory one. Operations that are identical or overlap are deduplicated. Duplication may happen because the ``operations'' in the implementation could contain placeholders to denote infinite sets of operations defined in \secref{def:4-op}. Say, two operations with placeholders represent the sets $O_1$ and $O_2$. $O_1$ duplicate $O_2$ if $O_1 \subseteq O_2$ or vice versa. In this case, we keep only $O_2$, as it is more general. An extra constraint is imposed for commands retrieved from the denylist entries: the command with the proposed arguments must be blocked by the denylist. For example, \codespan{cat Cargo.toml} will be excluded from the denylist in \autoref{lst:2-running_example} because the denylist do not block it at all. As we argued previously, such a command is not what the denylist aims to block. The LLM will be re-prompted for candidates who do not meet this requirement. \looseness=-1



On the example \codespan{cp} present in \autoref{tab:5-inout}, the LLM may proposed a candidate argument list \codespan{\$P /dev/stdout} with three candidate operations: \codespan{op\_copy("\$P", "/tmp/f1")}, \codespan{op\_read("\$P")}, and \codespan{op\_read("project.env")}. \codespan{op\_read("\$P")} which contains the placeholder \codespan{\$P} is strictly more general than \codespan{op\_read("project.env")}, so the latter is discarded. After this stage, two candidate operations, \codespan{op\_copy("\$P", "/tmp/f1")} and \codespan{op\_read("\$P")}, with the candidate argument list \codespan{\$P /dev/stdout} are proposed.\looseness=-1

\subsection{Execution-based Validation}
\label{sec:5-method_validation}

\begin{figure}[t]
    \centering
    \includegraphics{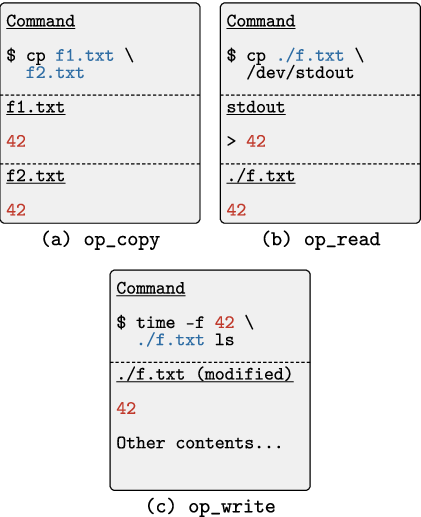}
    \caption{Execution-based validation. Each candidate is instantiated with a freshly generated flag file (in blue)/canary token (in red). The command is executed in a sandbox. A per-scope oracle inspects the pre- and post-execution states to determine whether the operation was actually performed.\looseness=-1}
    \label{fig:5-probes_and_goals}
\end{figure}

The validator decides, for each candidate operation, whether the command truly performs it when running on the host. Each of the seven scopes of \autoref{tab:4-scopes} is equipped with an \emph{oracle}. The command with arguments and the operation are instantiated by substituting the placeholders with freshly generated concrete values that serve as flag files or canary tokens (e.g., \codespan{\$P} to \codespan{./f.txt}). The command is then executed in a sandbox that records its side effects. If the operation is performed, the side effects must satisfy the oracles on the flag files and canary tokens. \autoref{fig:5-probes_and_goals} illustrates the oracles for three representative operations:\looseness=-1
\begin{packeditemize}
    \item \textbf{\codespan{op\_copy}.} The command is supplied with the source and target paths. The oracle passes if and only if the target exists after execution, and it contains the same content as the source path.\looseness=-1
    \item \textbf{\codespan{op\_write}.} The command is supplied with a flag file path already present in the host environment and a canary token to be written to the file. The oracle passes if and only if the canary appears in the file after execution.\looseness=-1
    \item \textbf{\codespan{op\_read}.} A flag file is seeded with a canary token and used to substitute for the command's file path placeholder(s). The oracle passes if and only if the canary appears in \codespan{stdout} or \codespan{stderr}.\looseness=-1
\end{packeditemize}
The remaining four operations in \autoref{tab:4-scopes} are validated by oracles of the same form: a freshly minted canary makes the command's effect observable, and an operation-specific oracle reports the presence or absence of that effect. Details are as follows:
\begin{packeditemize}
    \item \textbf{\texttt{op\_create(path)}.} \texttt{path} is instantiated with a fresh unoccupied path. The oracle passes if and only if a filesystem entry exists at \texttt{path} after execution.
    \item \textbf{\texttt{op\_delete(path)}.} A flag file is seeded at \texttt{path} before execution. The oracle passes iff \texttt{path} no longer exists afterward.\looseness=-1
    \item \textbf{\texttt{op\_chmode(path, mode)}.} A flag file is seeded with a known initial mode $m_0$ (e.g., \texttt{0644}) and \texttt{mode} is set to a target $m_1 \neq m_0$ (e.g., \texttt{0700}). The oracle passes iff the permission bits of \texttt{path} equal $m_1$ after execution. Choosing $m_1 \neq m_0$ prevents a no-op from passing.\looseness=-1
    \item \textbf{\texttt{op\_chowgr(path, owner\_and\_group)}.} A flag file at \texttt{path} is seeded with a known owner and group, and \texttt{owner\_and\_group} names a distinct owner/group that exists on the host. The oracle passes iff the owner and group of \texttt{path} match the target afterward. \looseness=-1
\end{packeditemize}

Let's continue the example input \codespan{cp \$P /dev/stdout} and candidate scopes \codespan{op\_copy("\$P", "/tmp/f1")} and \codespan{op\_read("\$P")}. First, \codespan{\$P} is substituted by a freshly created flag file \codespan{./f.txt} containing the canary token \codespan{42}. Then, the resulting side effects are checked against the oracles of the two candidate operations. The oracle for \codespan{op\_copy(path1, path2)} requires that \codespan{path2} exists and holds the same content as \codespan{path1}. It fails in the current case because \codespan{/tmp/f1} was not created. The oracle for \codespan{op\_read(path)} requires that the content of \codespan{path} is present in \codespan{stdout} or \codespan{stderr}, which is true in our case. Therefore, the candidate operation \codespan{op\_read("\$P")} is validated. \looseness=-1

Two design choices govern the validator:
\begin{packeditemize}
    \item \textbf{Execution over LLM judgment.} Validation runs the command and observes its effect rather than asking a second LLM to adjudicate. Because the candidate is itself generated by an LLM, it can exhibit the full range of LLM failure modes, such as hallucinations or outdated usage~\cite{zhang2025llm}, and an LLM judge would share these failure modes rather than correct them.\looseness=-1
    \item \textbf{Deliberately loose oracles.} Each oracle checks only the essence of its operation. For \codespan{scope\_read}, the essence is whether the seeded token reaches \codespan{stdout}/\codespan{stderr}. Everything orthogonal to it, such as exit codes, warnings on \codespan{stderr}, or incidental output, should be ignored. A command that emits a warning or returns a non-zero status while still leaking the file's contents genuinely performs the operation, and a stricter oracle would wrongly reject it.\looseness=-1
\end{packeditemize}

\subsection{Repair Loop}
\label{sec:5-method_repair}

\begin{figure}[t]
    \centering
    \includegraphics{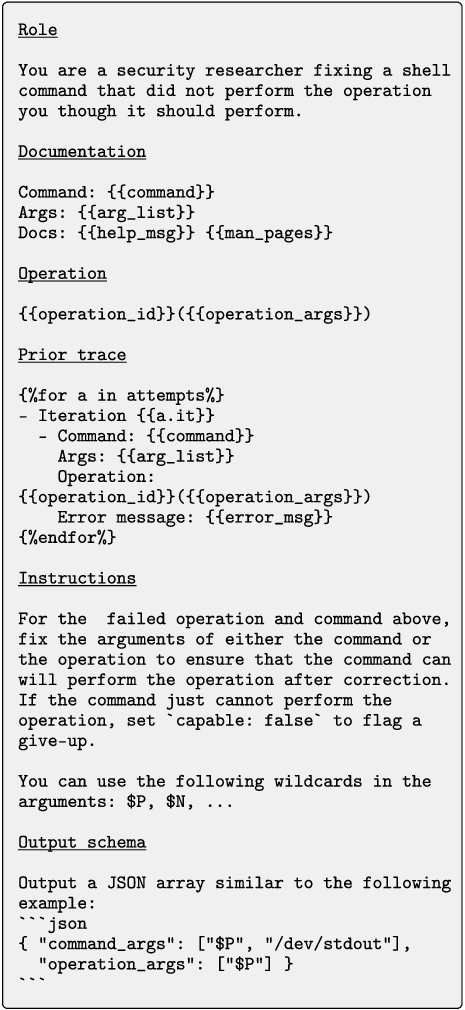}
    \caption{Prompt template used in the repair loop.}
    \label{fig:5-repair}
\end{figure}

A pair of a command and a candidate operation that fails validation, either because it raises errors or does not satisfy the operation's oracle, is returned to the LLM together with the captured \codespan{stdout}, \codespan{stderr}, the return code, and any validator-side diagnostics such as parsing errors. The LLM is asked to emit a single revised candidate that addresses the observed failure while still targeting the same operation, and the revision is sent through execution-based validation again. This loop repeats for at most $M$ (a configurable constant) rounds per original candidate and exits early as soon as a revision passes the oracle or once the LLM decides the candidate is not fixable. The repair loop's purpose is to recover \emph{near-misses}, such as unbalanced quotes or incorrectly used flags, which a validate-once pipeline would discard. For the example \codespan{cat \$N.env} in \autoref{tab:5-inout}, say, a candidate operation \codespan{op\_read(\$N.var)} fails because of the wrong suffix. In the repair loop, the LLM may fix it to a correct version \codespan{op\_read(\$N.env)} and succeed the re-run validation pass. The prompt used in the repair loop is in \autoref{fig:5-repair}. Note that we deliberately exclude the GTFOBins entry from the reference documentation, as it is community-compiled and more likely to introduce outdated or incorrect information. \looseness=-1

\subsection{Matching Denylists with Bypasses}
\label{sec:5-method_match}

The shared pipeline yields, for each command $c$, the operations it performs ($\mathcal{O}_h(c)$), and for each denylist, the operations it blocks ( $\mathcal{O}_h(L)$). For example, for $c$ being the command \codespan{cp}, it computes $\mathcal{O}_h(c)$ being $\{\texttt{op\_read("\$P")}, \texttt{op\_copy(\$P1, \$P2)}\}$, and for $L$ being the denylist in \autoref{lst:2-running_example}, it computes $\mathcal{O}_h(L)$ is $\{\texttt{op\_read("\$N.env")}\}$. Then, the two sets are simply intersected as $\mathcal{O}_h(c) \cap \mathcal{O}_h(L)$, and if the intersection is not empty, $c$ is a bypass of $L$. For each $o \in \mathcal{O}_h(c) \cap  \mathcal{O}_h(L)$, $c \succ_{h, L} L$. In our example, $\mathcal{O}_h(c) \cap \mathcal{O}_h(L) = \{\texttt{op\_read("\$N.env")}\}$, and thus \texttt{cp} is a bypass of \autoref{lst:2-running_example} that can read \texttt{.env} files. Note that in the intersection, the most concrete values that contain placeholders are preserved (e.g., \texttt{op\_read("\$N.env")} instead of \texttt{op\_read("\$P")} in the intersection). \looseness=-1

\section{Evaluation}
\label{sec:eval}

We collected a dataset of real-world command denylists used by terminal AI agents from GitHub. We applied \sysname on it to characterize the fragility of command denylists caused by incompletely blocked operations at scale and to draw insights for future practice and defenses.

\subsection{Research Questions}
\label{sec:eval-rqs}

The evaluation is framed against the formalization in \secref{sec:problem} and targets the following research questions: \looseness=-1
\begin{packeditemize}
    \item \textbf{RQ1 (Problem severity).} How severe is the command denylist fragility problem among the collected dataset? \looseness=-1
    \item \textbf{RQ2 (Severity in different cases).} Do factors such as how famous the project of a denylist is or what the agent is affect the severity of the problem? \looseness=-1
    \item \textbf{RQ3 (Root causes).} What are the root causes of incompletely blocked operations? \looseness=-1
    \item \textbf{RQ4 (Burden of fixes).} How difficult is it to repair a fragile denylist? Specifically, how many bypass commands must be added to fully block an operation? \looseness=-1
\end{packeditemize}

\subsection{Metrics}

We answer each RQ using metrics computed from the operations and bypasses identified by \sysname according to the formalization in \secref{sec:problem}: \looseness=-1
\begin{packeditemize}
    \item \textbf{RQ1.} Severity is measured by \emph{incompletely-blocked rate}: for each operation $o$, the fraction of denylists $L$ that incompletely block it among all that block it, i.e.,
    \[\frac{\left\Vert\left\{L \middle| o \blk_h^- L \right\}\right\Vert}{\left\Vert\left\{L \middle| o \blk_h L \right\}\right\Vert}\]
    where $\Vert\cdot\Vert$ means the size of a set.\looseness=-1
    \item \textbf{RQ2.} We compute the incompletely-blocking rate for each repository, and measure whether and how it varies among repositories with different numbers of GitHub stars and the agent the repositories use. \looseness=-1
    \item \textbf{RQ3.} We propose three hypotheses for the root causes of incompletely blocked operations, and compute the relations between the likelihood of a command becoming a bypass and the proxies of the hypothetical causes. \looseness=-1
    \item \textbf{RQ4.} The repair burden is proxied by the number of commands that the denylist must block in addition to cover all incompletely blocked operations. \looseness=-1
\end{packeditemize}

\subsection{Dataset Construction}
\label{sec:eval-dataset}

\begin{table}[t]
\centering
\begin{tabularx}{\linewidth}{Xrrrr}
\toprule
\textbf{Stars} & \textbf{Claude Code} & \textbf{Codex} & \textbf{OpenCode} & \textbf{Total} \\
\midrule
0--99      &  936 & 130 & 563 & 1629 \\
100--999   &   57 &   4 &   8 &   69 \\
$\ge$1000  &    7 &   2 &   2 &   11 \\
\midrule
Total      & 1,000 & 136 & 573 & 1709 \\
\bottomrule
\end{tabularx}
\caption{Stars of the GitHub repositories where the denylists are collected from.}
\label{tab:6-star}
\end{table}

The evaluation targets denylists for three widely used AI agents: Claude Code, Codex, and OpenCode. We collect the denylists from GitHub in their per-project setting files. Other AI agents, such as Copilot CLI~\cite{allowing} and Antigravity~\cite{google}, also follow the three-list gating architecture. However, their denylists are either set globally in the user's home directory rather than the project directory (and thus not uploaded to remote repositories) or specified on the command line rather than in a configuration file. It is worth noting that we did not include OpenClaw, as it was being developed at an extremely rapid pace (one release per day) at the time this paper was written, and its architecture and configuration schema are subject to drastic changes~\cite{2026openclaw}.\looseness=-1

\begin{figure}[t]
    \centering
    \includegraphics{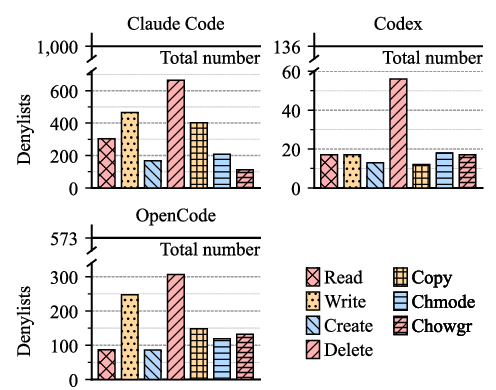}
    \caption{Number of denylists that block each operation, grouped by agent. The \codespan{op\_} prefix in the operation IDs is omitted.}
    \label{fig:6-scope_dist}
\end{figure}

\paragraph{Representativeness} The collected denylists well reflect how users use AI agents in the real world. In total, 24,453 repositories we find on GitHub contain denylists for Claude Code, 142 for Codex, and 1,253 for OpenCode. The number of repositories that contain Claude Code denylists is capped at 1000 via random sampling to limit the time required for the experiments. It is worth noting that the shell-command denylist is a rather new feature of Codex, marked as experimental~\cite{rules} at the time this paper is being written, so the number of collected denylists of Codex is relatively small. Before the feature was shipped, Codex lacked a gating mechanism for shell commands and relied solely on the AI model's decisions and a sandbox that restricted access by file path. \autoref{tab:6-star} shows the distributions of the numbers of GitHub stars these repositories receive, which typically have long tails: most have only a few stars, while a few receive many ($\ge$1,000). The repositories use 152 programming languages, and cover 1,643 topics (tagged by their maintainers). \autoref{fig:6-scope_dist} shows, for each of the seven operations in \autoref{tab:4-scopes}, the number of denylists that cover it (i.e., $p \in B_h(L)$). All operations are well represented, though with an imbalance across kinds that we attribute to users' preferences and expectations: \codespan{op\_delete}, for instance, is covered by the most denylists, which is intuitively explained by the destructive nature of deletion. Users tend to be more alert to such operations.\looseness=-1

\subsection{Experiment Settings and Costs}
\label{sec:eval-setup}

\paragraph{Host environments} All measurements are taken with respect to a reference host $h$: a Debian 13.5 Docker image with the default GNU userland, which provides the executables a developer running a terminal AI agent on a typical Linux laptop would have. \sysname instantiates the command universe $C_h$ of \cref{def:4-full} by walking the executable search directories. During execution-based validation (\secref{sec:5-method_validation}), each candidate is instantiated with freshly minted flag files and canary tokens in a per-run temporary directory within a sandbox using Bubblewrap~\cite{containers} and OverlayFS~\cite{overlay}. \looseness=-1

\paragraph{Models, costs, and time consumption.} The candidate enumeration and the iterative repair stages are driven by Claude Sonnet~4.6 with default settings. The repair loop is capped at 5 iterations per candidate, exiting early on the first revision that passes the oracle or when the model reports the candidate as unfixable. We access the model through AWS Bedrock and run the pipeline with 16 threads in parallel. The full evaluation consumed 3 wall-clock hours and \$25 in inference. \looseness=-1

\subsection{Results and Findings}
\label{sec:eval-results}

\subsubsection{RQ1 (Problem severity)}
\label{sec:eval-rq1}

RQ1 studies how often a blocked operation is left incompletely blocked. For each operation, we compute the incompletely blocked rate for each kind of operation.

\begin{figure}[t]
    \centering
    \includegraphics{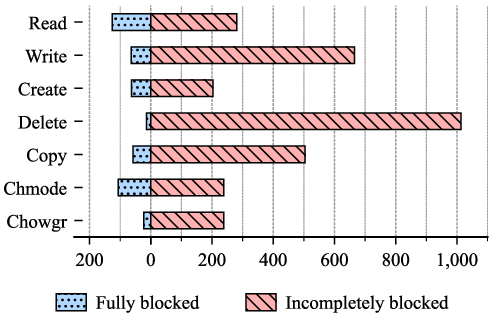}
    \caption{For each operation, the number of denylists that fully block it and incompletely block it. \looseness=-1}
    \label{fig:6-rq1_back2back}
\end{figure}

\autoref{fig:6-rq1_back2back} shows the number of denylists that incompletely or fully block an operation. Among the seven types of operations, denylists that incompletely block an operation account for the largest share (69.0--98.6\%). The result means that most denylists overlooked at least one bypass for the operation they target, invalidating their protection against it. \looseness=-1 

\vspace{1.5ex}

\noindent\fbox{\parbox{0.975\linewidth}{\textbf{Finding 1.} The problem of fragile command denylists caused by incompletely blocked operations is severe. 69.0--98.6\% of the denylists that target to block an operation overlook at least one bypass command for that operation.\looseness=-1}}

\subsubsection{RQ2 (Severity in different cases)}
\label{sec:eval-rq2}

\begin{figure}[t]
    \centering
    \begin{subfigure}[b]{\linewidth}
    \centering
    \includegraphics{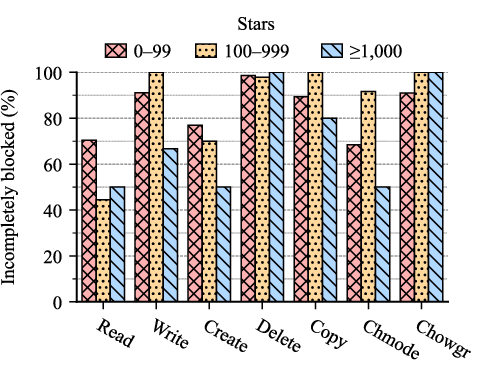}
    \caption{Incompletely blocked operations with respect to stars.}
    \label{fig:6-rq2_scope_incomplete_star}
    \end{subfigure} \\
    \begin{subfigure}[b]{\linewidth}
    \centering
    \includegraphics{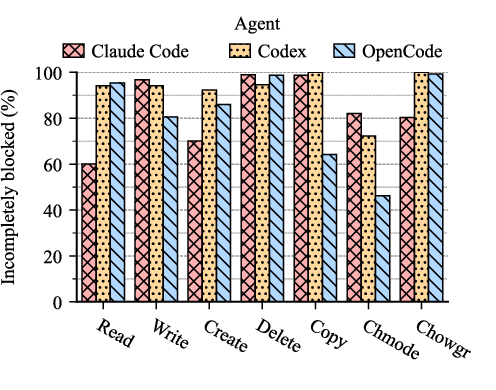}
    \caption{Incompletely blocked operations with respect to agents.}
    \label{fig:6-rq2_scope_incomplete_agent}
    \end{subfigure}
    \caption{Breakdown of incompletely blocked rate. \looseness=-1}
    \label{fig:6-rq2_scope_incomplete}
\end{figure}

RQ2 studies whether fragility is concentrated in particular kinds of projects or agents. \autoref{fig:6-rq2_scope_incomplete} reports, for each operation, the incompletely blocked rate, broken down by repository popularity (\autoref{fig:6-rq2_scope_incomplete_star}) and by agent (\autoref{fig:6-rq2_scope_incomplete_agent}).\looseness=-1

The incompletely-blocked rate stays high across all three star number tiers in \autoref{fig:6-rq2_scope_incomplete_star}, though in 4 out of the seven kinds of operations, projects with the most stars ($\ge 1,000$) have the lowest incompletely-blocked rate. However, even the lowest incomplete-blocked rate for the operations is still higher than 44.4\%. Similarly, the incomplete-blocked rate remains uniformly high across all three agents, consistently exceeding 46.2\%, and none shows a tendency to lower values.\looseness=-1

\vspace{1.5ex}

\noindent\fbox{\parbox{0.975\linewidth}{\textbf{Finding 2.} The fragile denylist problem is constantly severe across different projects and agents, as reflected by the uniformly high incomplete-blocked rates ($> 44.4\%$ across projects and $> 46.2\%$ across agents). Denylists from projects with more stars sometimes are less fragile, though.\looseness=-1}}


\subsubsection{RQ3 (Root causes)}
\label{sec:eval-rq3}

To find the root causes of incompletely blocked operations, we first propose several explanatory hypotheses and then validate them statistically. Note that the hypotheses are not meant to be a complete list but rather to provide initial insights and directions for future research. The hypotheses are as follows: \looseness=-1
\begin{packeditemize}
    \item \textbf{H1 (Ignorance).} The denylist authors do not know that the bypass commands exist. \looseness=-1
    \item \textbf{H2 (Versatility).} A command that performs several operations is deliberately left unlisted so that it remains usable for the operations the authors do \emph{not} wish to block, even though it may perform one that is blocked. \looseness=-1
    \item \textbf{H3 (Semantic gap).} The command's ability to perform the operation is hidden from its apparent purpose, so the author never associates it with the operation in the first place. \looseness=-1
\end{packeditemize}

\paragraph{H1 (Ignorance).} Intuitively, if the denylists overlook bypasses because their authors do not know that the bypass commands exist, then less popular (and thus less known) commands should be easier to become a bypass. To validate this hypothesis, we collect the rank of a command in the Debian Popularity Contest~\cite{debian} as a proxy for the popularity of the commands. Less popular commands correspond to larger rank values. Then, we apply equal-frequency binning~\cite{mining2006data} to split the commands into six bins based on their ranks and use a logistic regression model to estimate the frequency of bypass commands within each bin. The fitted model will predict, for a command at a given popularity rank, the probability that it is a bypass command for the operation.\looseness=-1

\begin{figure*}[t]
    \centering
    \includegraphics{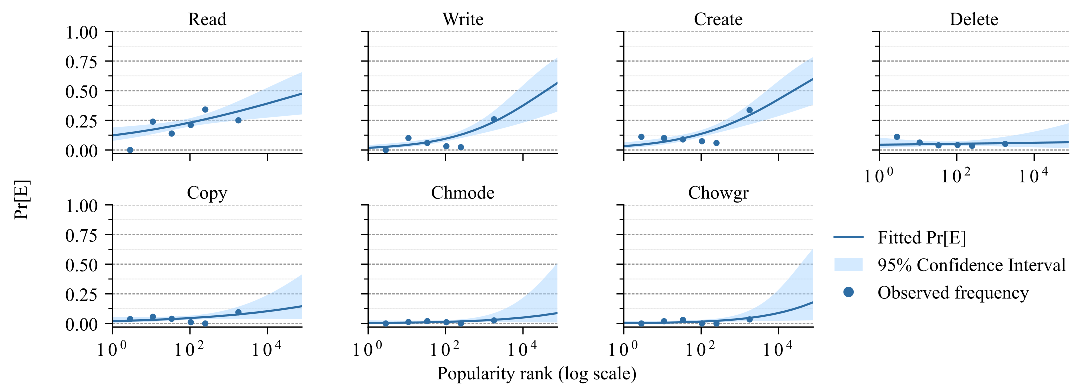}
    \caption{Fitted probability that a host command is a bypass for each operation as a function of its popularity rank (log scale), with 95\% CIs and empirical rates. $E$ is the event that a command $c$ becomes a bypass for any denylists ($c\succ_{h, o} L$ for any $L$), and $\mathrm{Pr}[\cdot]$ is the probability of an event.\looseness=-1}
    \label{fig:6-rq3_scope_logistic}
\end{figure*}

\autoref{fig:6-rq3_scope_logistic} shows the result. The lines are the fitted probability that a command bypasses the operation's denylist as a function of its popularity rank (on a log scale). The shadows are the corresponding $95\%$ confidence bands of these fitted curves. The points are the empirical bypass rates for the six equal-frequency bins, each plotted at its mean rank, and serve as a calibration reference for the fitted curve. The figure shows that for every operation except \codespan{op\_delete} the fitted probability rises with rank (the curves are especially steep for \codespan{op\_read}, \codespan{op\_write}, and \codespan{op\_create}), meaning that less popular (greater rank values) commands are markedly more likely to bypass the denylist, whereas under \textsf{Delete} popularity carries essentially no signal. This supports H1: the fitted probability of being a bypass rises with
rank for all seven operations, and the logistic regression across operations gives highly significant positive slopes (lowest $p$ being $1.02\times 10^{-6}$). Thus, less popular commands are markedly more likely to become bypasses.

\paragraph{H2 (Versatility).} Intuitively, if denylists overlook bypasses because multi-purpose commands are deliberately left unlisted to keep them usable for the operations the authors \emph{do} wish to permit, then a command that spans more operations should be more likely a bypass across the denylists. To validate this hypothesis, for each command $c$, we count the number of denylists it bypasses ($\left\Vert\left\{L \middle| c\succ_{h, o} L\right\}\right\Vert$), aggregated across all seven operations, and group the commands according to how many operations they perform: those performing more than one operation (multi-op) versus a single operation (single-op). We compare the two groups with a Mann-Whitney $U$ test~\cite{hollander2013nonparametric} and report two statistics. The $p$-value indicates whether the two groups' bypass counts differ, and Cliff's $\delta$ is an effect size in $[-1, 1]$ that captures how large the difference is and in which direction, with a positive value meaning multi-op commands bypass more denylists than single-op ones. Typically, a $p$-value below 0.05 indicates a statistically significant difference, and the magnitude of Cliff's $\delta$ is conventionally read as negligible below 0.147, small below 0.33, medium below 0.474, and large otherwise. \looseness=-1

\autoref{tab:any-denylist} shows the result. Multi-op commands bypass a median of 664 denylists, more than twice the 281 bypassed by single-op commands, and the gap is statistically significant (Mann-Whitney $U{=}2482$, $p{=}0.006$) with a non-negligible effect size (Cliff's $\delta{=}0.30$). This supports H2: a command that performs several operations remains available as a bypass for any of those operations that a given denylist does not cover, so the more operations a command spans, the more denylists it slips through.\looseness=-1

\paragraph{H3 (Semantic gap).} Intuitively, if denylists overlook bypasses because a command's ability to perform an operation is hidden behind its apparent purpose, then the overlooked bypasses should concentrate among commands that do not advertise the operation in their everyday documentation. To validate this hypothesis, we use the community-maintained TLDR pages~\cite{2026tldr-pages}, which list each command's most common usages, as a proxy for a command's apparent purpose. We rerun the \sysname pipeline using only the TL;DR documentation in the prompts to identify the obvious operations each command performs. Then, for each operation, we split the host commands into those whose TLDR page states the operation and those whose page does not, and compute the percentage of validated bypasses in each group. If a semantic gap were the dominant cause, bypasses should be more frequent among commands whose TLDRs do not mention the operation.\looseness=-1

\autoref{tab:6-rq3_bypass_tldr} shows the result, which runs opposite to what H3 predicts. For every operation, validated bypasses are far more common among commands whose TLDR states the operation than among commands whose TLDR does not, at 40.1\% versus 5.7\% overall and, for instance, 45.0\% versus 15.3\% for \codespan{op\_read}. The overlooked bypasses are therefore mostly commands whose TL;DR already states the operation, so their capabilities are advertised rather than hidden, and the semantic gap is at best a secondary cause. This rejects H3: the commands authors miss are documented in plain sight.\looseness=-1

\vspace{1.5ex}

\noindent\fbox{\parbox{0.975\linewidth}{\textbf{Finding 3.} Two causes of the fragility can be statistically confirmed. First, ignorance (H1): the authors of the denylists do not know about the less popular commands that become bypasses. Second, command versatility (H2): commands that perform several operations in bypass are deliberately unlisted for their benign uses, while these commands may be used as bypasses to perform the operations the denylists aim to block. \looseness=-1}}

\begin{table}[t]
\centering

\begin{tabularx}{\linewidth}{Xrr}
    \toprule
    \textbf{Command group} & $N$ & \textbf{Median [IQR]} \\
    \midrule
    Multi-op ($>\!1$)  & 35  & 664\,[280,\,1013] \\
    Single-op ($=\!1$) & 109 & 281\,[281,\,497]  \\
    \midrule
    Mann--Whitney $U$ & \multicolumn{2}{r}{2482}  \\
    $p$-value         & \multicolumn{2}{r}{0.006} \\
    Cliff's $\delta$  & \multicolumn{2}{r}{0.30}  \\
\bottomrule
\end{tabularx}
\caption{Denylists that a command can bypass, grouped by whether the command performs multiple operations or a single operation. $N$ is the number of commands in each group, and IRQ means interquartile range.}
\label{tab:any-denylist}
\end{table}

\begin{table}[t]
    \centering
    \begin{tabularx}{\linewidth}{lRRRRRRR}
    \toprule
         & \myhead{Read} & \myhead{Write} & \myhead{Create} & \myhead{Delete} & \myhead{Copy} & \myhead{Chmode} & \myhead{Chowgr}\\
    \midrule
         TLDR Op     & 45.0 & 23.7 & 44.6 & 35.1 & 75.0 & 40.0 & 40.0\\
         Other & 15.3 & 7.9 & 9.1 & 3.7 & 4.0 & 1.3 & 1.3 \\
    \bottomrule
    \end{tabularx}
    \caption{Percentage of host commands that are validated bypasses, split by whether the operation appears in the command's TLDR documentation. \emph{TLDR Op} covers commands whose TLDR page states the operation. \emph{Other} covers those whose does not.}
    \label{tab:6-rq3_bypass_tldr}
\end{table}

\subsubsection{RQ4 (Burden of fixes)}
\label{sec:eval-rq4}

RQ4 examines the cost of repairing a fragile denylist. Fully blocking an operation ($o \blk^{+}_{h} L$) requires the denylist to match \emph{every} host command that performs $o$ ($c \vdash_{h, a} o$. The repair burden may therefore be proxied by the number of bypass commands a defender must add to the denylist. \autoref{tab:6-rq4_bypass} reports the minimum, mean, and maximum number of distinct validated bypasses that must be added per operation. \looseness=-1

On the reference host, fully blocking \codespan{op\_read} takes, on average, 217 additional commands (up to 222), and closing every operation that a denylist blocks at once takes, on average, 180 commands (up to 603). Even the cheaper operations are far from a one-line fix: \codespan{op\_write} and \codespan{op\_create} average 97 and 98 bypasses each. The numbers are substantial for anyone who intends to fix the denylists. \looseness=-1

\vspace{1.5ex}

\noindent\fbox{\parbox{0.975\linewidth}{\paragraph{Finding 4.} Repairing a fragile denylist by enumerating bypasses requires adding numerous commands (217 on average for \codespan{op\_read}). \looseness=-1}}

\begin{table}[t]
    \centering
    \begin{tabularx}{\linewidth}{lRRRRRRR}
    \toprule
         & \myhead{Read} & \myhead{Write} & \myhead{Create} & \myhead{Delete} & \myhead{Copy} & \myhead{Chmode} & \myhead{Chowgr} \\
    \midrule
         Min  & 200 & 5   & 63  & 38 & 27 & 1  & 7  \\
         Mean & 217 & 97  & 98  & 54 & 39 & 3  & 11 \\
         Max  & 222 & 106 & 180 & 58 & 40 & 13 & 12 \\
    \bottomrule
    \end{tabularx}
    \caption{Min/mean/max number of bypass commands.}
    \label{tab:6-rq4_bypass}
\end{table}

\section{Related Work}
\label{sec:related}

\paragraph{Agent safety and tool-use sandboxing.} A growing line of work studies how to make AI agents safe when they are granted access to real tools. Proposals include sandboxed execution environments for agent actions~\cite{ruan2024identifying, wu2025isolategpt}, capability-scoped tool registries~\cite{shi2026progent, zhu2025miniscope, ji2026taming}, and human-in-the-loop approval workflows~\cite{weng2026what, wang2026reframing, lee2025takedown}. A parallel line of work replaces or augments the human approver with an LLM-based judge that inspects each proposed command in context~\cite{zheng2023judging}. Such judges sidestep approval fatigue, but they reason over the same command-string surface and inherit the hallucination and adversarial-suffix failure modes noted in \secref{sec:background}. \looseness=-1

\paragraph{LLM red-teaming and jailbreaks.} Red-team studies of LLMs have primarily targeted the model's own output, including prompt injection~\cite{hong2025sok, geng2026prompt}, jailbreaks that elicit prohibited content~\cite{wei2023jailbroken, zou2023universal}, and tool-misuse via indirect prompting~\cite{greshake2023not, zhan2024injecagent}. \looseness=-1

\paragraph{Living-off-the-land attacks.} The community databases GTFOBins~\cite{gtfobins} and LOLBAS~\cite{lolbas} catalogue standard binaries that can be abused for living-off-the-land attacks. Prior work has used these databases to study real intrusions~\cite{ongun2021living-off-the-land}. \looseness=-1
\section{Limitations and Future Work}
\label{sec:discussion}

\paragraph{Completeness of LLM-proposed bypasses.} \sysname{} discovers bypasses by prompting an LLM to enumerate candidate commands and operations, and an LLM-based generator can only recall a subset of the bypasses a host actually admits. Our measurements are therefore a \textit{lower bound} on denylist fragility: the incompletely-blocked rates and the repair burdens we report can only grow as enumeration improves, and the severity of the problem should be greater than what we have reported, which is already rather concerning. Crucially, this incompleteness affects only the magnitude of the effect, not its direction, considering the validation stage in our pipeline. Our study may encounter false negatives, but all the reported bypasses are true positives. \looseness=-1

\paragraph{Scope of operations.} We deliberately restrict the blocked operations under study to seven file-system operations and exclude others, including, most notably, network operations such as data exfiltration and reverse connections, to keep the evaluation tractable. File-system effects are durable and directly observable in the post-execution host state, which admits the small, robust oracles our validator relies on. In contrast, network effects would require modeling external endpoints and substantially heavier instrumentation. We leave it for future research to study the operations this paper didn't consider. \looseness=-1

\paragraph{Better command gating.} Our results indicate that the three-list command-gating mechanism itself is quite fragile, considering that the denylists overlook bypasses. However, this mechanism is prevalent due to the huge flexibility it provides. Future research and practice may investigate how to pair the three-list command gating with other techniques, such as the aforementioned capability-based sandbox~\cite{sandbox} and LLM auditor~\cite{claude}, to achieve a better trade-off between flexibility and security. \looseness=-1
\section{Conclusion}
\label{sec:conclusion}

This paper presents the first systematic characterization of command denylist fragility in terminal AI agents. It formalizes the operations that denylists aim to block and defines when a command can be a bypass to perform these operations. It proposes \sysname{}, an LLM-driven pipeline that automatically enumerates and validate bypasses for denylists. Applying \sysname{} to 1,709 real-world denylists (13,332 rules) collected from GitHub, we find that the denylist fragility problem is severe and pervasive: 69.0--98.6\% of the denylists that target an operation overlook at least one validated bypass. We also study and confirm several root causes of the problem. In addition, we find that fixing a fragile denylist will require completing the denylist to match numerous extra bypass commands, which makes the burden to fix these denylists heavy. These findings show that a static command-string denylist is inadequate as the load-bearing layer of agent command gating, and they motivate a shift toward better approaches that combine the list-based command gating mechanism with other techniques to achieve a better trade-off between flexibility and security for terminal AI agents. We hope our pipeline, dataset, and findings support future research and practice in this field. \looseness=-1

\clearpage
\section*{Ethics Considerations}
All the experiments, including those comparing different techniques on the bug-injected benchmarks, are conducted locally. No harm is done to real-world systems. Also, we have responsibly disclosed the new vulnerabilities (and bugs) to relevant stakeholders, including Anthropic and the project that uses the denylist in \autoref{lst:2-running_example}, and at this time of writing, the disclosure and possible countermeasure are still under review and consideration.

\bibliographystyle{IEEEtran}
\bibliography{references}

\end{document}